\documentclass[12pt]{article}
\usepackage{graphicx}
\usepackage{epsfig}
\usepackage{enumerate}

\newcommand{\beq}{\begin{equation}}
\newcommand{\eeq}{\end{equation}}
\newcommand{\beqa}{\begin{eqnarray}}
\newcommand{\eeqa}{\end{eqnarray}}
\newcommand{\beqar}{\begin{eqnarray*}}
\newcommand{\eeqar}{\end{eqnarray*}}

\newcommand{\norm}[1]{\raise.3ex\hbox{:}#1\raise.3ex\hbox{:}}

\newcommand{\p}{\Phi}

\begin{document}

\setlength{\unitlength}{1mm}

\thispagestyle{empty} \rightline{KIAS-P07013}

\vspace*{3cm}

\begin{center}
  {\bf \Large Stationary black holes and attractor mechanism}\\
  \vspace*{2cm}

  {\bf Dumitru Astefanesei}\footnote{E-mail: {\tt dumitru@th.phys.titech.ac.jp}}
  {\bf and Hossein Yavartanoo}\footnote{E-mail: {\tt yavar@phya.snu.ac.kr}; }

  \vspace*{0.2cm}

  {\it $^1$\sl Global Edge Institute, Tokyo Institute of Technology}\\
  {\it \sl Tokyo 152-8550, JAPAN}\\[.5em]
  {\it $^{2}$\sl Center for Theoretical Physics and BK-21 Frontier Physics Division
    Seoul National University, Seoul 151-747 KOREA}
    \vspace*{3mm } \\[.5em]

  \vspace{2cm} {\bf ABSTRACT} 
  \end{center}
We investigate the symmetries of the near horizon geometry of extremal stationary black
hole in four dimensional Einstein gravity coupled to abelian gauge fields and neutral
scalars.  Careful consideration of the equations of motion and the boundary conditions at
the horizon imply that the near horizon geometry has $SO(2,1)\times U(1)$ isometry. This
complements the rotating attractors proposal of hep-th/0606244 that had assumed the
presence of this isometry. The extremal solutions are classified into two families
differentiated by the presence or absence of an ergo-region. We also comment on the 
attractor mechanism of both branches.

\vfill \setcounter{page}{0} \setcounter{footnote}{0} \newpage

\section{Introduction}

The attractor mechanism plays a key role in understanding the entropy of
non-supersymmetric extremal black holes in string theory \cite{dst, astefanesei2}.
In certain cases, the macroscopic entropy of extremal non-supersymmetric attractor
horizons can be matched to the weak coupling statistical entropy despite the fact that
these quantities do no seem to be protected by supersymmetry \cite{emparan, 
eh, hlm, Maldacena:1996gb, Dabholkar:1997rk}.

It was originally noticed in \cite{{Bardeen:1999px}} that the extremal four dimensional Kerr
and Kerr-Newman black holes have an $SO(2,1)\times U(1)$ isometry.  The last year, the
authors of \cite{astefanesei},
found even more four dimensional extremal black holes had this isometry. Emboldened by
this observation, they found that, for four dimensional stationary extremal black holes, in a
theory of gravity with neutral scalar fields non-minimally coupled to abelian gauge
fields, one can generalise the entropy function formalism of \cite{Sen:2005wa}
simply by assuming an $SO(2,1)\times U(1)$ near horizon geometry. 

The generalised entropy function is constructed, on an $SO(2,1)\times U(1)$ symmetric background, by
taking the Legendre transform (with respect to the electric charges and angular momentum) of the 
reduced Lagrangian evaluated at the horizon.  Extremising the entropy function is equivalent to the equations
of motion and its extremal value corresponds to the entropy. Since the entropy function
depends only on the near horizon geometry, its extremum and hence the entropy is
independent of the asymptotic data. This is precisely the attractor behaviour. However, if
the entropy function has flat directions something interesting happens: while the extremum
remains fixed, flat directions will not be fixed by near horizon data and can depend on
the asymptotic moduli.

There exist two distinct
branches of stationary extremal black hole solutions which, in \cite{astefanesei},
are dubbed `ergo-' and `ergo-free' branches according to their properties.\footnote{The existence 
of two branches in the moduli space of extremal rotating black holes was discussed for the 
first time in \cite{rash}.} The first branch,
also known as the fast branch, can exist for angular momentum of magnitude larger than a
certain lower bound and does have an ergo-region. On the other hand, the ergo-free branch
can exist only for angular momentum of magnitude less than a certain upper bound. The
ergo-free branch can also be smoothly connected to a static extremal black hole.

The entropy function has no flat directions for the ergo-free branch: the scalar and all other
background fields at the horizon are independent of the asymptotic data. However, there is
a drastic change for the ergo-branch --- the entropy function has flat directions:
despite the entropy being independent of the moduli, the near horizon fields are dependent
on the asymptotic data. 

We find it significant that, the existence of an ergo-region allows energy to be extracted
classically either by the Penrose process for point particles or by superradiant
scattering for fields. It is tempting to believe that the presence of the ergo-sphere is
intimately related to the appearance of flat directions. One might say that the
ergo-branch, not completely isolated from its environment due to these processes, retains
some dependence on the asymptotic moduli.  From this perspective, it is amazing that the
black hole is isolated enough for the entropy to remain independent.\footnote{It is possible 
that the addition of higher derivative terms might lift these
  flat directions. It would be interesting to see whether this would erase the
  ergo-sphere.}

A consistent microscopic picture for Kaluza-Klein (KK) black hole in agreement with the
macroscopic analysis of rotating attractors \cite{astefanesei}
was provided in \cite{emparan, eh}.
That is, the D-brane model reproduces the entropy of KK black hole, while the
{\it mass gets renormalized} from weak to strong coupling {\it just} for the
{\it ergo-branch} black hole solutions in agreement with the existence of the flat
directions in the entropy function for this branch. Emparan and Maccarrone, \cite{emparan},
have also provided a microscopic interpretation for the superradiant ergosphere --- even
if the temperature is vanishing, the extremal black holes with ergosphere correspond to
states with both left- and right-moving excitations such that the open strings can combine
and the emission of closed strings is possible. The extraction of energy should reduce the
angular momentum in such a way that the event horizon area is increasing (it can not
decrease in the classical processes). Indeed, since the left-moving excitations have spin,
the emitted closed string will necessarily carry angular momentum away from the black
hole.

In this note we fill up a gap in the proposal of \cite{astefanesei}
by proving that the near horizon geometry of extremal rotating black holes in Einstein-Hillbert
gravity coupled to abelian gauge fields and neutral scalar fields has an
enhanced $SO(2,1)\times U(1)$ symmetry. Unlike the static case where the near horizon
geometry is $AdS_2\times S^2$, the $AdS_2$ part does not decouple from the angular part
and the values of the moduli at the horizon have an angular dependence. Also, by adding
angular momentum to static black holes, the $SO(3)$ symmetry of the sphere is broken to
$U(1)$. However, the near horizon geometry is still universal in the sense that is still
independent of the coupling constants and is determined just by charges and angular 
momentum parameter. The attractor
mechanism is related to the extremality rather than to the supersymmetry property of the
theory/solution. Indeed, the enhanced symmetry of the near horizon geometry and the long 
throat of $AdS_2$ is at the basis of the attractor mechanism for stationary black holes 
\cite{astefanesei, Sen:2005wa}.

\section{Generalities}

We consider a theory of gravity coupled to a set of masless scalars and vector fields, whose
general bosonic action has the form
\begin{eqnarray}
  \nonumber
  I[G_{\mu\nu},\phi^i,A_{\mu}^I]
  \!&=&\!
  \frac{1}{k^2}\int_{M} d^{4}x \sqrt{-G}[ R-2g_{ij}(\phi)\partial_\mu\phi^i\partial^\mu\phi^j-f_{AB}(\phi)F^{A}_{\mu\nu}F^{B\, \mu\nu} \\
  &&-\frac{1}{2\sqrt{-G}}{\tilde f}_{AB}(\phi) F^A_{\mu \nu}
  F^B_{\rho \sigma} \epsilon^{\mu \nu \rho \sigma} ]\, ,
  \label{actiongen}
\end{eqnarray}
where $F^A_{\mu\nu}$ with $A=(0, \cdots N)$ are the gauge fields, $\phi^i$ with
($i=1, \cdots, n$) are the scalar fields, and $k^2=16\pi G_4$. The moduli
determine the gauge coupling constants and $g_{ij}(\phi)$ is the metric in the moduli
space. We use Gaussian units to avoid extraneous factors of $4\pi$ in the gauge fields,
and the Newtons's constant is set to $G_4=1$.

Varying the action we obtain the following equations of motion for the metric, moduli, and
the gauge fields:

\begin{equation}
  R_{\mu\nu}-2g_{ij}\partial_{\mu}\phi^i\partial_{\nu}\phi^j
  =f_{AB}\left(2F^A_{\phantom{A}\mu\lambda}F^{B\phantom{\nu}\lambda}_{\phantom{B}\nu}-
    {\textstyle \frac{1}{2}}G_{\mu\nu}F^A_{\phantom{a}\alpha\lambda} F^{B \alpha\lambda} \right)
  \label{einstein}
\end{equation}
\begin{equation}
  \frac{1}{\sqrt{-G}}\partial_{\mu}(\sqrt{-G}g_{ij}\partial^{\mu}\phi^j)
  =\frac{1}{4} \frac{\partial f_{AB}}{\partial \phi^i} F^A _{\phantom{A}\mu\nu} F^{B\, \mu\nu}
  +\frac{1}{8\sqrt{-G}}\frac{\partial \tilde f_{AB}}{\partial \phi^i} 
  F^A_{\phantom{A}\mu\nu} F^B_{\phantom{B}\rho \sigma} \epsilon^{\mu\nu\rho\sigma} \\
  \label{dilaton}
\end{equation}

\begin{equation}
  \partial_{\mu}\left[\sqrt{-G}\left(f_{AB} F^{B\, \mu\nu}
      + \frac{1}{2\sqrt{-G}} {\tilde f}_{AB}F^B_{\phantom{B}\rho\sigma}
      \epsilon^{\mu\nu\rho\sigma}\right) \right] =  0.
  \label{gaugefield}
\end{equation}
To get the equations of motion, we have varied the moduli and the gauge fields
independently. The Bianchi identities for the gauge fields are
$F_{\phantom{A}\,[\mu\nu;\lambda]}^{A}=0$.

We are interested in stationary black hole solutions to the equations of motion. In
general relativity the boundary conditions are fixed. However, in string theory one can
obtain interesting situations by varying the asymptotic values of the moduli and so, in
general, the asymptotic moduli data should play an important role in characterizing these
solutions. Indeed, the non-extremal black hole solutions are characterized by the usual
conserved charges and also by the scalar charges --- the scalar charge is defined as the
monopole in the multipoles expansion of the scalar field at the boundary. Thus all its
properties are moduli dependent, e.g. the entropy depends by the asymptotic values of the
moduli. However, the entropy of extremal solutions obtained by taking the smooth limit
when the temperature is vanishing is independent of the asymptotic moduli data. We will
see in the next section that the enhanced symmetry of their near horizon geometry make
them special in this regard.

Now let us write down the most general stationary black hole solution by using just its
symmetries.\footnote{The thermodynamics of the non-extremal black hole solutions using the
  method developed in \cite{Astefanesei:2005ad, Mann:2005yr, Astefanesei:2006zd}
  will be presented in \cite{eu}.} An asymptotically flat spacetime is stationary if
and only if there exists a Killing vector field, $\xi$, that is time-like at spatial
infinity --- it can be normalized such that $\xi^2=-1$. It was also been shown that
{\it stationarity} implies {\it axisymmetry} \cite{Hollands} and so the event
horizon is a Killing horizon. Using the time-independence and axisymmetry we can write the
most general stationary metric with an `axial' Killing vector, ${\partial_\phi}$, as

\begin{equation}
  \label{metric} 
  ds^2=g_{tt}dt^2 +
  2g_{t\phi}dt\,d\phi+g_{\phi\phi}d\phi^2+g_{rr}dr^2+
  g_{\theta\theta}d\theta^2.
\end{equation}

The event horizon of a stationary black hole is a Killing horizon of
${\partial_t}+\omega{\partial_\phi}$, where the constant coefficient $\omega$ is the
angular velocity of the horizon. It is convenient to rewrite the metric (\ref{metric}) in
the ADM form
\begin{equation}
  ds^2=-N^2\,dt^2 +
  \gamma_{ij}\,(dx^i+N^i\,dt)(dx^j+N^j\,dt)=-N^2\,dt^2 +
  g_{\phi\phi}(d\phi+N^\phi dt)^2 +g_{rr}dr^2+g_{\theta\theta}d\theta^2,
\end{equation}
and so we obtain:
$$
N^2=\frac{(g_{t\phi})^2}{g_{\phi\phi}}-g_{tt}, \,\,\,\,
N^{\phi}=\frac{g_{t\phi}}{g_{\phi\phi}}, \,\,\,\, \gamma_{ij}=g_{ij}.
$$
The shift vector, $N^\phi$, evaluated at the horizon reproduces the angular velocity of the horizon:
$$
\label{angularvelocity} \omega=-\left.N^{\phi}\right\vert_H =
-\left.\frac{g_{t\phi}}{g_{\phi\phi}}\right\vert_H\,.
$$
By eliminating the conical singularity in the Euclidean $(\tau=it,r)$ sector, we obtain
the temperature
\beq
\label{temperature}
T=\frac{1}{\Delta\tau}=\left.\frac{(N^2)'}{4\pi\sqrt{N^2g_{rr}}}\right\vert_H.
\eeq

\section{Near horizon geometry of extremal black holes}
We consider a generic covariant two derivative gravity Lagrangian that has three basic
components: metric, scalars, and gauge fields. We show that, given a few simple
assumptions, the near horizon geometry of a stationary, extremal spinning black hole
solutions of this Lagrangian necessarily has the near
horizon symmetry $SO(2,1)\times U(1)$. To prove the previous statement, we make use of the following ingredients:\\
$\bullet$ Symmetries: we assume time independence and axisymmetry;\\
$\bullet$ The black hole is extremal --- in other words the surface gravity (temperature) is zero;\\
$\bullet$ We expand the fields near the horizon and take a scaling limit;\\
$\bullet$ Gauge choices;\\
$\bullet$ Finiteness of certain physical quantities;\\
$\bullet$ Equations of motion;\\
$\bullet$ Spherical topology of the horizon.

\subsection{Constraining the metric}
As a warm-up exercise, we begin by examining $4$-dimensional spherically symmetric black
holes by using the following ansatz:
\begin{eqnarray}
  ds^{2} & = & -a(r)^{2}dt^{2}+a(r)^{-2}dr^{2}+b(r)^{2}d\Omega^{2}.
  \label{metric2}
\end{eqnarray}
The near horizon geometry of the extremal black holes can be obtained in two steps: first, take the extremal
limit when the temperature is vanishing (this is a smooth limit on the Lorentzian section) and then obtain the near horizon 
geometry.
We expand the metric components near the outer horizon and for the non-extremal solution ($r_+\neq r_-$) we obtain:
\begin{equation}
  a^2=\rho f(r)=\rho(f_0+f_1\rho+f_2\rho^2+...), \,\,\,\,\,\,\,\,
  b^2=\frac{\rho(\rho+\epsilon)}{a^2}=\frac{\rho+\epsilon}{f_0+f_1\rho+f_2\rho^2+...}.
\end{equation}
and so the temperature is $f_0=4\pi T$.
Here we used a coordinates system such that the horizon is at $\rho=r-r_+=0$ and defined the non-extremality parameter 
$\epsilon=r_+-r_-$.

The extremal limit is obtained for $4\pi T=f_0 \rightarrow 0$ and to obtain the near horizon geometry we also take $\rho=0$. 
By changing the coordinate $\tau=t/f_1$ one can easily obtain the $AdS_2\times S^2$ explicitly
\begin{equation}
  ds^2=\frac{1}{f_1}(-\rho^2d\tau^2+\frac{1}{\rho^2}d\rho^2)+\frac{1}{f_1}d\Omega^2.
\end{equation}

We use a similar method to obtain the near horizon geometry of stationary extremal black holes. However, the extremal limit in 
this case is more subtle since we should also consider the non-diagonal component ($\sim d\phi dt$) of the metric. Let us first 
rewrite the metric components in a more useful form:
\beq 
N^2=(r-r_+)(r-r_-)\mu(r,\theta), \,\,\,\,\,\,
N^{\phi}=-\omega+(r-r_+)\eta(r,\theta), \,\,\,\,\,\,
g_{rr}=\frac{1}{(r-r_+)(r-r_-)\Lambda(r,\theta)}, 
\eeq 
where
$\mu(r,\theta), \eta(r, \theta),$ and $\Lambda(r, \theta)$ are regular functions. The temperature can be read off
as before and using eq.~(\ref{temperature}) we obtain
\beq
T=\frac{r_+-r_-}{4\pi}\sqrt{\mu(\theta)\Lambda(\theta)},
\eeq
where $\mu(\theta)$ and $\Lambda(\theta)$ are the values of $\mu(r,\theta)$ and
$\Lambda(r,\theta)$ at the outer horizon. For a non-extremal black hole the temperature
(surface gravity) is finite and constant on the horizon and so we obtain
$\sqrt{\mu(\theta)\Lambda(\theta)}=C$, where $C$ is a constant that depends on the charges
$P,Q$ and the (mass and angular) parameters $m,a$.  Expanding the ADM form of the metric 
near the outer horizon we obtain the following metric:
\begin{equation}
  -(r-r_+)(r-r_-)\mu(\theta)dt^2+g_{\phi\phi}[d\phi+(r-r_+)\eta(\theta)dt]^2+
  \frac{1}{(r-r_+)(r-r_-)\Lambda(\theta)}dr^2+g_{\theta\theta}d\theta^2.
\end{equation}

To obtain the near horizon geometry, we first construct the following family of metrics
\begin{equation} r\rightarrow r_{+} +\lambda r , \;\;\;\;\;\;\;\;\; t \rightarrow
  \frac{t}{\lambda},
\end{equation}
where $\lambda$ is an arbitrary parameter. There is a smooth limit $\lambda \rightarrow 0$
for which the near horizon geometry is obtained. Obviously, this is important for
stationary field configurations where there exist also terms of the form $drdt$. This
limit is especially useful when we consider the near horizon expansion of the gauge
fields.

Taking the extremal limit ($r_+\rightarrow r_-$) and choosing a particular gauge we obtain
\begin{equation}
\label{important}
  ds^2=\mu(\theta)
  \left(
    -r^2dt^2 +
    \frac{dr^2}{C^2r^2}
  \right)
  +\frac{\sin^2\theta}{\mu(\theta)}(d\tilde{\phi}+r\eta(\theta)dt)^2+
  \frac{\mu(\theta)}{C^2}d\theta^2,
\end{equation}
where $\tilde{\phi} = \phi -\frac{\omega t }{\lambda}$. To obtain the above expression we use
the appropriate coordinates system  in which $g_{\theta\theta} = r^2 g_{rr}$ and the gauge
freedom to write $\mu(\theta) g_{\phi\phi}(\theta) = \sin^2\theta$. 
Here, we have considered the metric in a rotating frame with respect to a distant observer
with the angular velocity equal to that of the black hole. For a horizon with spherical
topology, we require 
\begin{equation}
  \frac{\sin^2\theta}{C^2\mu^2(\theta)} \sim 
  \left\{\begin{array}{cc}
      \theta^2 & \quad \theta\rightarrow0\\
      (\pi-\theta)^2 & \quad \theta\rightarrow\pi\\
    \end{array}
  \right.
\end{equation}
such that the deformed horizon, labelled by the coordinates $(\theta,\phi)$, is a smooth
deformation of the sphere. Unlike the static case, the fields at the horizon have an
angular dependence and so solving the attractor equations requires boundary conditions,
i.e. the values of the fields at the poles of the horizon.

Let us end up this subsection with an important comment about the extremal limit. For a
stationary black hole there are three intensive parameters associated to the horizon: the
angular velocity, the temperature, and the electric (magnetic) potential.  Thus, there are
two interesting extremal limits $T=0$ when the angular velocity is or is not vanishing.  
In the discussion section we comment further on the physics of the extremal black holes.

\subsection{Constraining the scalars and gauge fields}
\label{sigata}
Let us start by investigating the scalar and the gauge fields configuration in the near horizon limit. For simplicity, we do not 
carry on the moduli and gauge fields indices in this subsection --- we specialize to one scalar and one 
gauge field configuration, but the generalization to a configuration with more than one scalar and one 
gauge field is straightforward. Expanding the scalars at the horizon, 
$r=0$, we obtain 
  \begin{equation}
    \phi(r,t) = r^\alpha(\phi(\theta)+r\phi_1(\theta)+{\cal O}(r^2)+\ldots).
  \end{equation}
Requiring that the scalars are finite at the horizon, implies $\alpha\geq0$ and then by taking the 
scaling limit, $r\rightarrow\lambda r$, $\lambda\rightarrow0$, we find
  \begin{equation}
    \phi = 
    \left\{\begin{array}{cc}
        \phi_0(\theta) & \quad \alpha=0\\
        0 &\quad \alpha>0
      \end{array}
    \right.
  \end{equation}
in the near horizon region. We assume that the near horizon effective gauge coupling $f(\phi(\theta))$ is well behaved
and can be Taylor expanded around the poles, i.e.
\begin{equation}
  f(\phi(\theta))=f_0+f_1\theta+{\cal O}(\theta^2)
\end{equation}
Let us turn to the gauge fields and perform a similar analysis. We impose that the gauge fields are time-independent 
and start with the following ansatz 
  \begin{equation}
    A = A_t(r,\theta)dt+A_r(r,\theta)dr+A_\theta(r,\theta)d\theta+A_\phi(r,\theta)d\phi\, ,
  \end{equation}
that can be further simplified by choosing an appropriate gauge choice to fix $A_\theta=0$ (or $A_r=0$).
We can expand the gauge fields about the horizon as
  \begin{equation}
    A = r^\alpha\left[a_{t}(\theta)+{\cal O}(r)\right]rdt
    +r^\beta\left[a_{r}(\theta)+{\cal O}(r)\right]\frac{dr}{r}
    +r^\gamma\left[a_{\phi}(\theta)+{\cal O}(r)\right]d\phi
  \end{equation}
  Requiring $F^2$ remains finite at the horizon implies $\alpha,\beta,\gamma\geq0$.  We
  take $\alpha,\beta,\gamma=0$ so that after taking the scaling limit
  $r\rightarrow\lambda r$, $t\rightarrow t/\lambda$, $\lambda\rightarrow0$ we obtain a
  non-zero result. With this assumption the scaling limit gives
  \begin{equation}
    A = a_{t}(\theta)rdt
    +a_{r}(\theta)\frac{dr}{r}
    +a_{\phi}(\theta)d\phi+{\cal O}(\lambda)
  \end{equation}

The Einstein equations can be written as
\begin{equation}
  R_{\mu\nu}-2\partial_{\mu}\phi\partial_{\nu}\phi
  =f\left(2F_{\mu\lambda}F^{\phantom{\nu}\lambda}_{\nu}-
    {\textstyle \frac{1}{2}}g_{\mu\nu}F_{\alpha\lambda} F^{\alpha\lambda} \right)  
\end{equation}
The ($r\theta$) equation plays an important role in what follows: using the results from the appendix and the 
fact that $\zeta=0$, we get
\begin{equation}
  \frac{\sin^2(\theta)}{\mu^2(\theta)}\eta(\theta)\eta'(\theta)=0
\end{equation}
which implies $\eta(\theta)$ is, in fact, a constant. 
This was the last step in our proof --- it is straightforward to check that the metric (\ref{important})
with $\eta(\theta)$ a constant function has the $SO(2,1)\times U(1)$ isometry. 

\section{Attractor mechanism}
\label{moduli}

In this section, we consider the attractor mechanism for static and stationary black
holes. For the static black hole solutions, we show the equivalence of the entropy
function formalism and the effective potential method. Entropy function formalism 
was generalized to stationary black holes in \cite{astefanesei}. We comment on 
the role played by the enhanced symmetry of the near horizon geometry in decoupling 
the moduli equations of motion at the horizon from the bulk.

\subsection{Static black holes}
The Bianchi identity and equation of motion for the gauge fields can be solved by a field
strength of the form
\begin{equation}
  \label{fstrenghtgen}
  F^A=f^{AB}(Q_{B}-{\tilde f}_{BC}P^C) {1\over b^2} dt\wedge dr + P^A \sin \theta  d\theta \wedge d\phi,
\end{equation}
where $P^A, Q_{A}$ are constants that determine the magnetic and electric charges carried
by the gauge field $F^A$, and $f^{AB}$ is the inverse of $f_{AB}$.

As discussed in \cite{Goldstein:2005hq}, the equations of motion for the moduli are 
\begin{equation}
  \label{eq3}
  \partial_{r}(a^{2}b^{2}g_{ij}\partial_{r}\phi^j)=\frac{1}{2b^{2}}
  \frac{\partial V_{eff}}{\partial \phi^i}\, ,
\end{equation}
where $V_{eff}(\phi^i)$ is a function of scalars fields $\phi^i$ given by
\begin{equation}
  \label{defpotgen}
  V_{eff}(\phi_i)=f^{AB}(Q_A-{\tilde f}_{AC}P^C)(Q_B - {\tilde f}_{BD}P^D)+f_{AB}P^AP^B.
\end{equation}
It is clear from the equation (\ref{eq3}) that $V_{eff}(\phi^i)$ is an `effective potential' for the
scalar fields --- it plays an important role in describing the attractor mechanism
\cite{Goldstein:2005hq, Ferrara}. 

For the attractor mechanism it is sufficient for two conditions to be met.  First, for
fixed charges, as a function of the moduli, $V_{eff}$ must have a critical point. Denoting
the critical values for the scalars as $\phi^i=\phi^i_0$ we have,
\begin{equation}
  \label{critical}
  \partial_iV_{eff}(\phi_{i0})=0.
\end{equation}
Second, the matrix of second derivatives of the potential at the critical point,
\begin{equation}
  \label{massmatrix}
  M_{ij}={1\over 2} \partial_i\partial_jV_{eff}(\phi_{i0})
\end{equation}
should have positive eigenvalues.

Once the two conditions mentioned above are met it was argued in \cite{Goldstein:2005hq}
that the attractor mechanism works and the entropy is given by the effective potential at
the horizon.

The near horizon geometry is $AdS_2\times S^2$ and so we can apply Sen's entropy function
\cite{Sen:2005wa} to investigate the attractor behaviour of static extremal solutions. All
other background fields respect the $SO(2,1)\times SO(3)$ symmetry of
$AdS_2\times S^2$. We keep the analysis general in order to understand the role of
$V_{eff}$.

In \cite{Sen:2005wa}, Sen found that the entropy of a spherically symmetric extremal black
hole is the Legendre transform of the Lagrangian density --- the only requirements are
gauge and general coordinate invariance of the action. In fact, this is similar with a 
generalization of the Wald's formalism for extremal black holes and it is based on the 
observation that there is a {\it smooth} extremal limit on the Lorentzian section of a 
charged black hole.

The entropy function is defined as
\begin{equation}
  F(\overrightarrow{u},\overrightarrow{v},\overrightarrow{e},\overrightarrow{p})=
  2\pi(e_iq_i-f(\overrightarrow{u},\overrightarrow{v},\overrightarrow{e},\overrightarrow{p}))=
  2\pi(e_iq_i-\int d\theta d\phi\sqrt{-G}L),
\end{equation}
where $d\phi\sqrt{-G}L$ is the Lagrangian density, $q_i=\partial f/\partial e_i$ are the
electric charges, $u_s$ are the moduli values at the horizon, ${p_i}$ and ${e_i}$ are the
near horizon radial magnetic and electric fields, and $v_1$, $v_2$ are the sizes of
$AdS_2$ and $S^2$, respectively. Thus, $F/2\pi$ is the Legendre transform of $f$ with respect to the
variables $e_i$. Then, for an extremal black hole of electric charge $\overrightarrow{Q}$
and magnetic charge $\overrightarrow{P}$, Sen have shown that the equations determining
$\overrightarrow{u},\overrightarrow{v}$ and $\overrightarrow{e}$ are given by:
\begin{equation}
  \frac{\partial F}{\partial u_s}=0\,, \qquad \frac{\partial F}{\partial v_i}=0\,,
  \qquad \frac{\partial F}{\partial e_i}=0\,.
  \label{attractor}
\end{equation}
Thus, the black hole entropy is given by
$S=F(\overrightarrow{u},\overrightarrow{v},\overrightarrow{e},\overrightarrow{p})$ at the
extremum (\ref{attractor}). We observe that the entropy function,
$F(\overrightarrow{u},\overrightarrow{v},\overrightarrow{e},\overrightarrow{p})$,
determines the sizes $v_1$, $v_2$ of $AdS_2$ and $S_2$, and also the near horizon values
of moduli ${u_s}$ and gauge field strengths ${e_i}$.

Now, we are ready to apply this method to our action (\ref{actiongen}). The general metric
of $AdS_2\times S^2$ can be written as
\begin{equation}
  ds^2=v_1(-\rho^2d\tau^2+\frac{1}{\rho^2}d\rho^2)+v_2(d\theta^2+\sin^2\theta d\phi^2)
\end{equation}

The field strength ansatz is
\begin{equation}
  F^A=F^A_{r\tau}\,dr\wedge d\tau+P^A\sin\theta\,d\theta \wedge d\phi=e^A\,dr\wedge d\tau+P^A\sin\theta\,d\theta \wedge d\phi
\end{equation}
and so
\begin{eqnarray}
  && F(v_1,v_2,e,q,p)=2\pi [q_Ae^A-f(v_1,v_2,e,p)]\\
  \nonumber
  && f(v_1,v_2,e,p)=\frac{8\pi}{k^2}\left[-v_2+v_1-
    f_{AB}\left(\frac{-v_2}{v_1}e^Ae^B+\frac{v_1}{v_2}p^Ap^B\right)-2\tilde f_{AB}e^Ae^B\right]
\end{eqnarray}
The attractor equations are:
\begin{eqnarray}
  \label{at1}
  \frac{\partial F}{\partial v_1} & = & 0\,\,\,\Rightarrow \,\,\,1-\frac{v_2}{v_1^2}f_{AB}e^Ae^B-
  \frac{1}{v_2}f_{AB}p^Ap^B=0\\
  \label{at2}
  \frac{\partial F}{\partial v_2} & = & 0\,\,\,\Rightarrow \,\,\,-1+\frac{1}{v_1}f_{AB}e^Ae^B-
  \frac{v_1}{v^2_2}f_{AB}p^Ap^B=0\\
  \label{at3}
  \frac{\partial F}{\partial \phi^i} & = & 0\,\,\,\Rightarrow \,\,\,\frac{\partial f_{AB}}{\partial \phi^i}\left( p^Ap^B-e^Ae^B \right) = 2\frac{\tilde f_{AB}}{\partial \phi^i}e^Ap^B\\
  \label{at4}
  \frac{\partial F}{\partial e^A} & = & 0\,\,\,\Rightarrow \,\,\, q_A=\frac{16\pi}{k^2}\left(\frac{v_2}{v_1}f_{AB}e^B-\tilde f_{AB}p^B \right)
\end{eqnarray}
By combining the first two equations we obtain $v=v_1=v_2=f_{AB}(e^Ae^B+p^Ap^B)$ that is
expecting also from our near horizon geometry analysis above. It's also easy to check that
the entropy is given by $F$ at the attractor critical point:
\begin{equation}
  S=F=\frac{16\pi^2}{k^2}f_{AB}(e^Ae^B+p^Ap^B)=\pi v
\end{equation}
Using the electromagnetic field ansatz,(\ref{fstrenghtgen}), it can be easily shown that
$S=\pi V_{eff}$, $q_A=-Q_A$ (the $-$ sign appears because of our convention for
$F^A_{tr}$), and (\ref{at4}) matches the critical point condition of $V_{eff}$.

\subsection{Stationary black holes}
We have shown in the previous section that the near horizon geometry of extremal spinning
black holes has the symmetries of $AdS_2\times S^1$ and can be written as \beq
\label{eg1}
ds^2\equiv g_{\mu\nu}dx^\mu dx^\nu = v_1(\theta) \left(-r^2 dt^2+{dr^2\over r^2}\right) +
\beta^2 \, d\theta^2+ \beta^2\, v_2(\theta) (d\phi - \alpha r dt)^2 \eeq and the most
general field configuration consistent with the $SO(2,1)\times U(1)$ symmetry of
$AdS_2\times S^1$ is of the form: \beqa
&&\Phi^i =u^i(\theta) \nonumber \\
&& {1\over 2}\, F^{A}_{\mu\nu}dx^\mu\wedge dx^\nu = (e^A-\alpha b^A(\theta)) dr \wedge dt
+ \partial_\theta b^A(\theta) d\theta\wedge (d\phi - \alpha r dt)\, , \eeqa where
$\alpha$, $\beta$ and $e_i$ are constants, and $v_1$, $v_2$, $u^i$, and $b^A$ are
functions of $\theta$.  Here $\phi$ is a periodic coordinate with period $2\pi$ and
$\theta$ takes value in the range $0\le\theta\le \pi$.

Based on this observation, a generalized entropy function was proposed in  \cite{astefanesei}
\beq 
F \equiv 2\pi (J\alpha+Q_Ae^A-\int d\theta d\phi\sqrt{-detG}L)\, ,
\eeq 
and so there is one more attractor equation associated to the angular momentum $J$. Thus, the 
entropy and the near horizon background of a spinning extremal black hole are obtained by 
extremaizing this entropy function that depends only on the parameters labellig the near horizon 
background and the electric and magnetic charges and the angular momentum carried by the 
black hole. 

Interestingly, in all known cases, the appearance of flat directions in the entropy is
associated with the presence of an ergo-sphere. Since not all moduli are fixed at the horizon
the mass is not guaranted to be fixed. The microscopic analysis of \cite{emparan} confirms 
that in fact the mass is not fixed and so there is a nice microscopic interpretation for the 
ergo-branch. On the other hand, the slowly spinning extremal black holes in the ergo-free 
branch lack the rotational superradiance but can produce superradiant amplification of KK electric 
charged waves. However, this phenomenon can not be easily seen in the CFT since it is related 
to a modification of the central charge. 

One important question is if there is a similar effective potential for stationary black holes and if one 
can use a similar analysis as in the static case to study the attractor mechanism. Unfortunately, at 
this point, we have just shown that the equations of motion at the horizon decouple from the 
bulk ---- we hope to report a detailed analysis elsewhere. Here, let us just indicate the main step 
 in this analysis. We start by trying to solve the equations of motion and Bianchi identities for the 
gauge fields. However, unlike the static case,  we can not obtain a general expression for the gauge 
fields, but rather their expressions in terms of two unknown functions (A and u):
\begin{eqnarray}
&& F^M_{\; rt}= \partial_r A^M_t(r,\theta) \hspace{3cm} F^M_{\theta t}= \partial_{\theta} A^M_t(r,\theta) \cr\cr
&& F^{M\; r\phi} = f^{MN}(\phi^i)\; \frac{\partial_{\theta}u_N(r,\theta)}{\sqrt{-g}}  \hspace{12mm} F^{M\; \theta\phi} = -f^{MN}(\phi^i)\; \frac{\partial_r u_N(r,\theta)}{\sqrt{-g}}
\end{eqnarray} 
One can write down the expressions of the gauge fields in some concrete examples. However, an interesting exercise
is to work with this 
general form of the gauge fields and try to extract as much information as possible from the equations of motion. 
For the moduli the equations of motion become
\begin{eqnarray}
 \frac{1}{\sqrt{-G}}\partial_{\mu}(\sqrt{-G}\partial^{\mu}\phi^i) = 
\frac{1}{2} \frac{\partial f_{MN}}{\partial \phi^i}   \left ( \partial_r A^M \partial_r A^N + r^2  \partial_{\theta} A^M\partial_{\theta} A^N  \right) \cr\cr    +  \frac{1}{2} \frac{\partial f^{MN}}{\partial \phi^i} \left( \partial_r u_M \partial_r u_N+ r^2  \partial_{\theta} u_M \partial_{\theta} u_N \right)
 \end{eqnarray}
However, in this case, the moduli have also an angular dependence and the equations do not decouple. In principle 
one should be able to read off the effective potential from the right hand side of this equation, but that is not 
straightforward in this case --- an effective potenial for constant scalar fields was proposed in \cite{astefanesei}.
The best thing we can do is to check what is happening in the near horizon limit. After some tedious manipulations 
we found that in the near horizon limit the moduli equations are decoupled from the bulk. The scalar fields at the 
horizon have also an angular dependence and we obtain a system of distributions rather than functions. Thus, the 
boundary conditions, i.e. the values of the fields at the poles of the horizon, are important and the equations are 
difficult to be solved in a general case --- concrete examples are presented in \cite{astefanesei}. 

\section{Discussion}
Recently, after the proposal of Sen \cite{Sen:2005wa}, there was a lot of work on attractor 
mechanism and entropy function (see, e.g., \cite{multe}).
Motivated by the generalization of the attractor mechanism to non-supersymmetric
extremal stationary black holes, we investigated the near horizon geometry of spinning extremal
black holes in a theory of gravity with neutral scalar fields non-minimally coupled to abelian
gauge fields. We found that the near horizon geometry of these black holes has the
symmetry of $AdS_2\times S^1$ --- the $AdS_2$ part does not decouple from the 
angular part. Consequently, the horizons are attractors for the moduli and
their geometry is independent of the boundary moduli data.  One subtlety 
is that the extremal spinning black holes  are further divided in two branches: ergo- and ergo-free branch,
respectively.  In both cases the SO(2,1) isometry of $AdS_2$ is generated by the Killing
vectors: 
\beq L_1=\partial_t, \qquad L_0=t\partial_t-r\partial_r, \qquad
L_{-1}=(1/2)(1/r^2+t^2)\partial_t - (tr)\partial_r + (\eta /r)
\partial_\phi\,  ,
\eeq 
but they have distinct properties. The former is characterized by
an entropy function with flat directions and for the latter there is no flat
directions of the entropy function. If there are no flat directions then, clearly, the entropy is
independent of the moduli. On the other hand, if there are flat directions, then the
extremization of the entropy function does not determine all the moduli values at the horizon. 
Location of these parameters along the
flat directions may depend on the asymptotic values of the moduli. But since the entropy 
function  does not depend on the flat directions, the entropy is still independent of the asymptotic 
values of the moduli, and so has an attractor behaviour.

Let us comment now on the physics of the two branches. In general, in the supergravity
approximation, the entropy is a function of the duality invariant combinations
$D(Q_A, P^B)$, $S=\sqrt{\pm (|D|-J^2)}$ and the mass saturates an extremality bound that is
independent of the angular momentum parameter, $J^2$ --- the plus sign corresponds to the ergo-free 
branch and the minus sign to the ergo-branch. When $D=J^2$ the extremal horizon
disappears and becomes a naked singularity --- this situation resembles the static case
with one charge. Except this situation, the extremal limit has finite area and zero
surface gravity. The fastly spinning extremal black holes have a non-zero horizon angular
velocity and so their causal structure is  similar with Kerr
solution. Let us start with a non-extremal black hole that Hawking radiates.  Clearly,
Hawking radiation carries away the angular momentum and so the black hole is slowing
down. If the black hole is radiating away all the angular momentum before reaching the
extremal limit, then the corresponding solution will be in the ergo-free branch. On the
other hand, if the black hole reaches the extremal limit and the angular velociy is
non-zero, then there is radiation due to the ergo-region. If the evaporating process is
fine tuned such that the extremal limit is reached when $J^2=|D|$, then the black hole
behaves more as an elementary particle \cite{hw} --- there are potential
barriers outside the horizon which increase without bound.

In this paper we also tried to extend the analysis of the effective potential to
extremal spinning black holes. We have not been able to conclusively construct
an {\it explicit} effective potential, mainly because of technical obstacles. 
In the static case one can explictly check that te moduli are fixed at the attractor horizon that 
is a critical point for the effective potential. A similar analysis is difficult for the stationary case. 
However, by studying the equations of motion for the moduli, we concluded that they  
decouple from the bulk at the horizon.\footnote{This is 
not a sufficient condition for the attractor mechanism to exist. However, a rigorous proof was given in 
\cite{astefanesei} by using the entropy function formalism.} A complete determination of the scalar 
fields at the horizon needs also imposing the boundary conditions which are the values of the fields 
at the poles of the horizon.

The near horizon geometry of a stationary extremal black hole is universal and so the entropy does 
not depend of couplings. The extremality condition is very powerfull to force an attractor behaviour of the
horizon --- it is independent of the supersymmetry of the theory/solution. This does not come as a surprise, 
though, since  the near horizon geometry has an enhanced symmetry and the long throat of 
$AdS_2$ is the main ingredient for the existence of the attractor mechanism.

\section*{Acknowledgements}
We thank Kevin Goldstein for collaboration in the initial stages of this work and for further discussions.
It is also a pleasure to thank Soo-Jong Rey, Ashoke Sen, and Sandip Trivedi for useful conversations.
DA would like to thank KIAS, Seoul for hospitality during part of this work. DA has presented 
this work at ISM06 Puri (December 2006), KIAS, Seoul (February 2007), YITP, Kyoto (February, 2007), 
TITECH, Tokyo (May 2007) and he likes to thank the audience at all these places for their positive feedback. 
The work of DA has been done with support from MEXT's program ``Promotion of Environmental 
Improvement for Independence of Young Researchers'' under the Special Coordination Funds for 
Promoting Science and Technology, Japan. DA also acknowledges support from NSERC of Canada. 
HY would like to thank  the Korea Research Foundation Leading Scientist Grant (R02-2004-000-10150-0) 
and Star Faculty Grant (KRF-2005-084-C00003).
While this paper was being completed, ref.~\cite{reall} appeared which overlaps with the material 
presented in section 3.

\appendix
\section{The Maxwell equations in the near horizon limit}

In this apendix we explicitly obtain the equations of motion for the gauge fields in the near horizon 
limit. These expressions are useful in subsection (\ref{sigata}) --- for simplicity, we specialize again 
to a configuration with one scalar and one gauge field, but the generalization is straightforward.

The non-zero components of the Maxwell tensor are given by
\begin{equation}
  \left.\begin{array}{ccc}
      F_{t\mathrm{i}}&=&(a_t(\theta),ra_t'(\theta),0)\\
      F_{\mathrm{i}\theta} &=&(a_r'(\theta)/r,0,a_\phi'(\theta))  
    \end{array}
  \right\}
  \quad \mbox{where i}\in \{r,\theta,\phi\} 
\end{equation}
Raising the indices we obtain
\begin{eqnarray}
  F^{t\mathrm{i}}&=&
  -\frac{C^2}{\mu^2(\theta)}
  \left(a_t(\theta),\frac{\zeta(\theta)}{r},0\right)\\
  F^{\mathrm{i}\theta} &=&\frac{C^2}{\mu^2(\theta)}\left(
    C^2ra_r'(\theta),0,\frac{\mu^2(\theta)}{\sin^2\theta}a_\phi'(\theta)
    +\eta(\theta)\zeta(\theta)\right)\\
  F^{\phi r}&=& \frac{C^2}{\mu^2(\theta)}\eta(\theta) ra_t(\theta)
\end{eqnarray}
where $\zeta(\theta)=a_t'(\theta)-\eta(\theta) a_\phi'(\theta)$.

Maxwell's equations are
\begin{equation}
  \p_\mu(\sqrt{-G}f F^{\mu\nu})=0
\end{equation}

From the $r$-component of Maxwell's equation we obtain
\begin{equation}
  \p_\theta(\mu^{-1}(\theta)\sin(\theta)f(\theta)a_r'(\theta))=0
\end{equation}
which can be integrated to give
\begin{equation}
  a_r'(\theta)=\kappa_1\frac{\mu(\theta)}{f\sin\theta}\sim \frac{\kappa_1}{\theta}
\end{equation}
where we have assumed that the effective gauge coupling at the north pole, $f(\theta=0)$,
is well behaved. Now, for $F^2$ to be finite at $\theta=0$, we require $\kappa_1=0$,
i.e. $a_r'=0$. This in turn means that $a_r(\theta)$ does not contribute to the Maxwell
tensor and can be gauged away.

Similarly from the $t$-component of Maxwell's equation we obtain
\begin{equation}
  \p_\theta(\mu^{-1}\sin(\theta)f(\theta)\zeta(\theta))=0
\end{equation}
which, by an argument similar to the one for $a_r'$ above, implies $\zeta$ is zero.
Some important relations used in this derivation are
\begin{equation}
  \sqrt{-G}=C^{-2}\mu(\theta)\sin\theta
\end{equation}
and
\begin{eqnarray}
  (\p_s)^2&=& g^{\mu\nu}\p{_\mu}\p{_\nu}\nonumber\\
  &=&\frac{1}{\mu(\theta)}
  \left(-\left[{r^{-1}}\p_t-\eta\p_\phi\right]^2
    +C^2r^2\p^2_{r}
    +C^2\p^2_{\theta}
  \right) 
  +\frac{\mu(\theta)}{\sin^2\theta}\p^2_{\phi}
\end{eqnarray}
\begin{equation}
  F^2=\frac{2C^2}{\mu^2}\left(-a_t^2-\zeta^2+C^2a_r^2\right)+\frac{2C^2}{\sin^2\theta}(a_\phi')^2
\end{equation}

\end{document}